\documentclass[aps,prl,twocolumn,superscriptaddress,amsmath]{revtex4-1}
\usepackage[pdftex]{graphicx,color}
\usepackage{dcolumn}
\usepackage{bm}

\usepackage{amsfonts}
\usepackage{amssymb}
\usepackage{amsmath}
\usepackage{amsthm}

\usepackage{subfigure}
\usepackage{rotate}

\usepackage[pdftex]{hyperref}

\let\oldmarginpar\marginpar
\renewcommand\marginpar[1]{\-\oldmarginpar[\raggedleft\tiny #1]%
{\raggedright\tiny #1}}

\DeclareMathOperator{\Tr}{Tr}

\newcommand{\bra}[1]{\langle#1|}
\newcommand{\ket}[1]{|#1\rangle}

\newcommand{\f}[2]{{\ensuremath{\mathchoice%
       {\dfrac{#1}{#2}}
       {\dfrac{#1}{#2}}
       {\frac{#1}{#2}}
       {\frac{#1}{#2}}
       }}}

\graphicspath{{figs/}}

\begin{document}


\title{Finite size scaling bounds on many-body localized phase transitions}

\author{A. Chandran}
\affiliation{Perimeter Institute for Theoretical Physics, Waterloo, Ontario N2L 2Y5, Canada}   \email{achandran@perimeterinstitute.ca}

\author{C. R. Laumann}
\affiliation{Department of Physics, University of Washington, Seattle, WA 98195, USA}
\affiliation{Perimeter Institute for Theoretical Physics, Waterloo, Ontario N2L 2Y5, Canada} 

\author{V. Oganesyan}
\affiliation{Department of Engineering Science and Physics,
College of Staten Island, CUNY, Staten Island, NY 10314, USA}
\affiliation{Initiative for the Theoretical Sciences, The Graduate Center, CUNY, New York, NY 10016, USA}

\date{\today}

\begin{abstract}
Quantum phase transitions are usually observed in ground states of correlated systems.
Remarkably, eigenstate phase transitions can also occur at finite energy density in disordered, isolated quantum systems. 
Such transitions fall outside the framework of statistical mechanics as they involve the breakdown of ergodicity. 
Here, we consider what general constraints can be imposed on the nature of eigenstate transitions due to the presence of disorder.
We derive Harris-type bounds on the finite-size scaling exponents of the mean entanglement entropy and level statistics at the many-body localization phase transition using several different arguments.
Our results are at odds with recent small-size numerics, for which we estimate the crossover scales beyond which the Harris bound must hold.
\end{abstract}

\maketitle

Progress in the study of interacting localization has brought to light new \emph{eigenstate phase transitions} in the highly excited states of disordered isolated quantum systems \cite{Pal:2010gr, PhysRevX.4.011052, Huse:2013aa, Vosk:2014fv, Kjall:2014aa, Chandran:2014kq, Bahri:2013ux, Khemani:2015fj, Potter:2015qd}. 
Across an eigenstate phase transition, properties of individual many-body eigenstates exhibit singularities.
Such transitions may be invisible in the traditional equilibrium statistical ensembles, especially when they involve an extreme breakdown of ergodicity as in the many-body localized (MBL) phase \cite{Pal:2010gr, PhysRevX.4.011052, Huse:2013aa, Vosk:2014fv, Kjall:2014aa, Chandran:2014kq, Bahri:2013ux, Khemani:2015fj, Potter:2015qd,Anderson:1958p7531,Gornyi:2005fv, Basko:2006aa,Oganesyan:2007aa,Znidaric:2008aa, Monthus:2010vn,Berkelbach:2010aa,Aleiner:2010ir,Bardarson:2012gc,Bauer:2013jw,DeLuca:2013ba,Swingle:2013it,Serbyn:2013he,Serbyn:2013cl,Huse:2014xe,Laumann:2014aa,Chandran:2014aa,Imbrie:2014jk,Grover:2014aa,Ros:2015rw,PhysRevLett.114.160401,Mondragon-Shem:2015cs,Torres-Herrera:2015qe,Tang:2015th}. 
Many-body localized phases have recently been experimentally observed in cold atomic and trapped ion systems \cite{Schreiber:2015aa,Smith:2015aa,Bordia:2015aa}.
The general theory of eigenstate transitions is largely unknown as the usual framework of the equilibrium critical theory need not apply.
Thus, it is useful to find general constraints to guide their study.

The most dramatic putative example of an eigenstate phase transition is the delocalization transition from a localized phase into a thermal phase satisfying the eigenstate thermalization hypothesis (ETH) \cite{Deutsch:1991ss, Srednicki:1994dw, Rigol:2008bh}. 
This signals the restoration of local thermalization at the level of individual eigenstates. 
To date, both numerical and phenomenological studies of delocalization in one dimension have found this kind of transition \cite{Pal:2010gr,PhysRevLett.114.160401, Kjall:2014aa,Luitz:2015fj,Vosk:2014aa, Potter:2015ab, Zhao:2015aa}.

A widely used order parameter for this transition is the disorder averaged entanglement entropy $[S]$ in eigenstates, which shifts from an area law in the localized phase to a volume law in the thermal \cite{Bauer:2013jw}.
Assuming that the transition is continuous and that the entanglement entropy density $[s]$ of a small fraction of the system satisfies a finite-size scaling ansatz,
\begin{align}
\label{eq:sscaling}
[s](L,\delta) \sim \frac{1}{L^a} \tilde{s}(L^{1/\nu} \delta)
\end{align}
we show that $a=0$ and $\nu \ge 2/d$ where $d$ is the spatial dimension. We thus prove a Harris bound for the mean entanglement entropy density at the MBL-ETH transition. This complements Grover's recent result\cite{Grover:2014aa} on the critical value of the scaling function.
The elementary arguments we present also apply directly to the mean level statistics ratio $[r]$, another commonly used diagnostic \cite{Oganesyan:2007aa}.

The original Harris criterion that $\nu \ge 2/d$ concerned the stability of clean equilibrium transitions to the introduction of disorder \cite{Harris:1974aa}.
Chayes, Chayes, Fisher and Spencer (CCFS) generalized this exponent bound to arbitrary disorder-driven transitions by defining a finite-size length $\xi_{FS}$ associated with the tails of the distribution of the order parameter \cite{Chayes:1989aa,PhysRevLett.57.2999}. 
When a traditional coarse-graining picture holds, central limiting behavior governs the distribution of the order parameter away from the transition. We then expect $\xi_{FS}$ to coincide with the correlation length governing the mean.
This is the reason the CCFS result is often interpreted as a proof of the Harris bound.

This argument could be applied to the entanglement entropy of the localized phase in the vicinity of a delocalization transition in $d\geq 2$.
Crudely, the entanglement entropy of a sub-region comes from independent blocks of size $\xi_{loc}$ straddling the boundary of the region, where $\xi_{loc}$ is a localization length.
Thus, for sub-regions significantly bigger than $\xi_{loc}$, the entanglement entropy shows central limiting behavior.
Consequently, the finite size scaling of the mean must be controlled by $\xi_{FS}$ and obey the Harris bound.

While this argument is appealing, it relies on several assumptions. First, it does not apply to non-spatial order parameters such as the level statistics ratio. Second, it doesn't apply to $d=1$ where most studies of the transition take place. Finally, it ignores the possibility of Griffiths type effects that could spoil the central-limiting behavior. For example, the mean susceptibility in the paramagnetic phase of the random transverse field Ising model functions as an order parameter for the transition into the Griffiths phase. However, it does not have central limiting behavior and violates the Harris bound. See also examples in Ref.~\cite{Pazmandi:1997aa}.

The mean scaling bounds we derive in this article are independent of these considerations. In particular, we can derive Harris bounds in any $d$ for $[s]$ and $[r]$ with no reference to detailed probability distributions.

In the following, we first derive an elementary lemma regarding the scaling of the means of bounded random variables. This immediately leads to an exponent bound $\nu \ge 2/(d+2a)$ for any bounded order parameter with a scaling form as in Eq.~\eqref{eq:sscaling}. For completeness, we recap the result of CCFS and the associated definition of $\xi_{FS}$. 
We then catalogue the behavior of several common order parameters used in the study of eigenstate phase transitions and discuss the relevance of these bounds to current numerical studies.
We conclude with some open questions for future study.

\vspace{2mm}

\emph{Lemma---} Consider a collection of $L^d$ sites subject to independent random disorder fields (or couplings) $h$ with probability $p_\lambda(h)$ dependent on disorder parameter $\lambda$. For any bounded random variable $Y \in [-1,1]$,
\begin{align}
\label{Eq:Lemma}
	\left|\f{d[Y]}{d\lambda}\right| \le \alpha L^{d/2}
\end{align}
where $\alpha^2 = \int dh\,\f{1}{p_\lambda(h)} \left(\f{dp_\lambda(h)}{d \lambda}\right)^2$ is an $O(L^0)$ constant and $[\cdot]$ denotes averaging with respect to the disorder.

\emph{Proof---}
The lemma is a direct generalization of the proofs presented in \cite{PhysRevLett.57.2999,Chayes:1989aa}. For simplicity, we review the argument for the case of a discrete bimodal distribution $p_\lambda(h) = \lambda \delta(h-h_1) + (1-\lambda)\delta(h-h_2)$. In this case, the probability of a disorder realization may be simply expressed $\Pr[\Omega] = \lambda^{n_1}(1-\lambda)^{n_2}$ where $n_i(\Omega)$ is the number of sites with field value $h_i$ in realization $\Omega$. We differentiate
\begin{align}
	[Y] &= \sum_{\Omega} \Pr[\Omega]\, Y[\Omega]
\end{align}
with respect to $\lambda$ to obtain,
\begin{align}
	\f{d[Y]}{d\lambda} &= \sum_{\Omega} \Pr[\Omega]\left(\f{n_1}{\lambda}-\f{n_2}{1-\lambda}\right)\,  Y[\Omega].
\end{align}
Taking the absolute value, recalling that $|Y| \le 1$ and applying the Cauchy-Schwarz inequality, 
\begin{align}
	\left|\f{d[Y]}{d\lambda}\right| &\le \sqrt{\sum_{\Omega} \Pr[\Omega] \left(\f{n_1}{\lambda}-\f{n_2}{1-\lambda}\right)^2}
\end{align}
The radicand is simply a joint moment of $n_1$ and $n_2 = L^d - n_1$. Thus,
\begin{align}
	\left|\f{d[Y]}{d\lambda}\right| &\le \alpha L^{d/2}
\end{align}
with $\alpha =  \sqrt{\lambda^{-1}+(1-\lambda)^{-1}}$ as required. 
The generalization to discrete distributions with $q>2$ disorder values is straightforward while that to continuum distributions ($q\to\infty$) requires somewhat more care \cite{Chayes:1989aa}.
QED.

Taking $Y$ to be an order parameter $X$, whose mean diagnoses a transition at $\delta = \lambda - \lambda_c = 0$, we immediately obtain the following theorem. 

\emph{\bf Mean Theorem ---} Any bounded random variable $X$ which satisfies the finite size scaling ansatz \footnote{We note that an $O(1)$ analytic background for $[X]$ does not change the bound.},
\begin{align}
\label{eq:fssansatz}
[X](L,\delta) \sim \f{1}{L^a} \tilde{X}(L^{1/\nu} \delta)
\end{align}
near a critical point at $\delta = 0$ has a scaling exponent 
\begin{align}
\label{eq:weakbound}
\nu \ge \f{2}{d+2a}
\end{align}

This bound is not as strong as the usual Harris bound $\nu \ge 2/d$ in the presence of a non-zero scaling dimension $a$. 
However, it assumes only the commonly used finite size scaling hypothesis, while the following theorem, due to CCFS \cite{PhysRevLett.57.2999}, requires a different hypothesis:

\emph{\bf Tail Theorem (CCFS)---} Consider any random variable $X$ which satisfies the following two conditions with fixed $O(1)$ constants $a$ and $c>0$:
\begin{itemize}
\item At the critical point at $\delta = 0$: $\Pr[X>a] \ge 2 c $ for all $L$ large enough.
\item Away from the critical point at $\delta > 0$: there exists a maximum length $\xi_{FS}(\delta)$, such that $\Pr[X>a] < c$ for all $L > \xi_{FS}$.
\end{itemize}
Then, the finite-size scaling length is bounded below,
\begin{align}
\label{Eq:CCFSxi}
\xi_{FS} \geq \beta \delta^{-2/d}
\end{align}
where $\beta=(c/\alpha)^{2/d}$ is an $O(1)$ smooth function of $\delta$.   

\vspace{2mm}

\emph{Applications---}
Using the above results, we constrain the behavior of commonly used order parameters for several different eigenstate phase transitions.

\emph{Entanglement Entropy---} The entanglement entropy $S$ of a subregion $A$ of linear dimension $L_A$ within an eigenstate $\ket{E_n}$ is defined by
\begin{align}
	S(L,L_A) = - \Tr \rho_A \log \rho_A
\end{align}
where $\rho_A = \Tr_{\bar{A}} \ket{E_n} \bra{E_n}$.
In the thermodynamic limit $L \to \infty$, the entanglement entropy density $s = S/L_A^d$ functions as an order parameter for the delocalization transition: it scales to zero in the localized phase and a finite value in the delocalized as $L_A \to \infty$.

In disordered systems, the spectrum itself is random and so one needs a prescription for choosing the eigenstates of interest near an energy $E$. 
One choice is simply to take the eigenstate $\ket{E_n}$ with the closest energy to $E$. Another possibility is to average over a subextensive energy window around $E$. The results below regarding the entanglement entropy $S$ are indifferent to this choice.

The entanglement entropy density $s$ is bounded between $0$ and $\log q$ where $q$ is the local Hilbert space dimension. 
Assuming a finite-size scaling ansatz for $[s]$ near a transition at disorder parameter $\delta = 0$, requires a two-parameter scaling form,
\begin{align}
\label{eq:twoparams}
  [s](L, L_A, \delta) \sim \frac{1}{L^a} \tilde{s}\left(L^{1/\nu} \delta, \frac{L_A}{L}\right)
\end{align}
where both $L^{1/\nu}\delta$ and $L_A/L$ are held fixed as $L \to \infty$. 
The scaling ansatz allows for the exponents and scaling functions to depend on other system parameters \footnote{In the usual renormalization group framework, this would arise in the presence of a line or surface of fixed points.}.
For example, they could depend on the energy density $e=E/L^d$, which we have also held fixed in this limit. 

At small $L_A/L$, we expect to access the scaling behavior in the thermodynamic limit at $L_A/L=0$.
In the localized phase, the correction due to fixed $L_A/L$ should be exponentially small in $L$ as the contributions to entanglement come from the boundary itself.
In the delocalized phase $[S]$ remains extensive although, in principle, the density $[s]$ could be reduced at non-zero $L_A/L$ from its thermodynamic value.
This does not happen in phases satisfying ETH, as the correction to $[s]$ (due to Page \cite{Page:1993zf}) vanishes as $1/L^d$.
The argument below only requires $[s]$ to be non-zero on the delocalized side and does not use this precise value. 
Thus from here on, we assume that $L_A/L$ is held fixed to some small fraction as the thermodynamic limit is approached.

There are two basic scenarios for the delocalization transition depending on whether the delocalized phase satisfies ETH. 

\emph{Localized-Delocalized (ETH)---}
In the ETH phase, the eigenstate ensemble reproduces the statistical ensemble for all local measurements \cite{Deutsch:1991ss, Srednicki:1994dw, Rigol:2008bh}. 
In particular, the entanglement entropy density $[s]$ is the thermal entropy density.
This implies that $[s]$ jumps at the transition $\delta = 0$.
The finite-size scaling hypothesis requires that $a = 0$ as a scaling dimension $a>0$ would impose a continuous turn on $[s] \sim \delta^{\nu a}$. 
Thus, by the Mean Theorem, $[s]$ must exhibit scaling with an exponent satisfying the Harris bound $\nu \ge 2/d$.

In $d\geq 2$, we argued before that the probability distribution of $s$ in the localized phase is likely Gaussian. 
If this is the case, the CCFS length coincides with the finite-size scaling length controlling the mean.
In $d=1$, as less is known about the distribution of $s$, we merely speculate that the CCFS length can be constructed. 
Although it would also satisfy the Harris bound, it need not coincide with the finite-size scaling length controlling the mean.

\emph{Localized-Delocalized (non-ETH)---}
If the delocalized phase fails to satisfy ETH, there are no further general constraints on the value of $[s]$. Consequently, the Mean Theorem provides a weaker bound as the presence of a non-trivial scaling dimension $a$ is not ruled out. 
If the CCFS length exists, it clearly satisfies the stronger Harris bound.

Unlike the MBL-ETH transition, there are several controlled examples of the localized-delocalized (non-ETH) transition. We review two as they illustrate the finite-size scaling behavior for $[s]$. However, both examples build on underlying critical theories with traditional coarse graining descriptions and thus the mean scaling behavior is constrained by CCFS.

The first example is the Anderson model wherein free fermions hop in a disordered background potential in $d=3$ \cite{Anderson:1958p7531}. 
As a function of decreasing disorder strength, this model undergoes a transition from a localized phase to a delocalized phase at infinite temperature. 
The delocalized phase is non-ETH as a finite density of the single-particle wavefunctions are localized at any non-zero disorder strength.
The value of $[s] \approx \rho \log 2$ where $\rho$ is the fraction of delocalized single particle states. 

The infinite temperature eigenstates are Slater determinants obtained by filling each single-particle orbital with probability one-half. 
Their entanglement structure follows directly from the spatial properties of the single-particle wavefunctions \cite{Peschel:1999aa,Jia:2008aa}. 
In particular, Eq.~\eqref{eq:twoparams} should hold with the single-particle localization length $\xi$ as the diverging length scale $\delta^{-\nu}$. 
The $d=3$ localization length exponent $\nu \approx 1.5 > 2/d$ so that the Harris bound is satisfied \cite{Kramer:1999gl, Evers:2008p10136}. 
Note however that the scaling dimension $a$ is non-zero at the transition because the extended states are multifractal and have vanishing density. 
The scaling dimension can be estimated by assuming that the extended states contribute order one and the critical states contribute $\sim 1/L_A^b$ to $[s]$. 
Using the density of extended states on the delocalized side $\rho \sim \delta^{1/2}$ and the density of critical states $\sim L_A^{-1/\nu}$, we obtain $1/\nu + b = a $ and $a=1/2\nu \approx 0.3$. 

The second example is the random Clifford circuit model in $d\geq 2$ (see Ref.~\cite{Chandran:2015aa} for details).
This is a tractable Floquet qubit model in which the localization-delocalization transition maps directly onto site percolation on the underlying lattice. 
Similar to the Anderson model, the delocalized phase possesses a finite density of fully localized conserved quantities which reduce the entanglement entropy density from $\log 2$. 
In each disorder realization, $S$ in a region $A$ is approximately determined by the number of sites within $A$ which are connected to the boundary. 
As finite-size scaling holds near the percolation transition, Eq.~\eqref{eq:twoparams} holds with the percolation length as the diverging length scale. 
The associated $\nu=\nu_{perc}$ satisfies the Harris bound in all dimensions $d\geq 2$ \cite{Shante:1971aa, Isichenko:1992fk}.
The scaling dimension $a$ is determined by the fractal dimension of the infinite cluster.

Finally, we note that both examples demonstrate that the two-parameter scaling at fixed $L_A/L$ accesses the same physics as that of the intrinsic limit $L_A/L =0$.

\vspace{2mm}

\emph{Level spacing ratio---}
The level spacing ratio $r$ of three consecutive energy levels $E_{n-1}< E_n < E_{n+1}$ is defined as:
\begin{align}
r_n \equiv \textrm{min}\left( \frac{\Delta E_n}{\Delta E_{n+1}}, \frac{\Delta E_{n+1}}{\Delta E_{n}} \right)
\end{align}
where $\Delta E_i = E_i - E_{i-1}$. The prescriptions described in the previous section can be used to define $r(E)$ near an energy of interest $E$.

The disorder distribution of $r$ distinguishes the ETH phase from localized phases \cite{Oganesyan:2007aa,Atas:2013aa}. 
In the ETH phase, the levels repel, exhibiting GOE statistics and $[r] \approx 0.527$. 
In the localized phase, the levels exhibit Poisson statistics and $[r]\approx 0.386$. The value of $[r]$ is an order parameter for the MBL-ETH transition which jumps across the transition. The value at the transition itself is under contention; recent work \cite{Serbyn:2015aa} suggests a third intermediate value determined by semi-Poisson statistics. 

Clearly, $r$ is a bounded random variable ($0<r<1$). Assuming that $[r]$ obeys a finite-size scaling ansatz of the form in Eq.~\eqref{eq:fssansatz} at the MBL-ETH transition, it immediately follows that $a=0$ as $[r]$ jumps at the transition. Thus, $\nu \geq 2/d$ by the Mean Theorem.

The divergence of the length scale $\xi_{FS}$ associated with the tails is more subtle as $\xi_{FS}$ may not exist. 
In particular, if the order parameter $X$ is chosen to be $r_n$ where $E_n$ is the closest energy to $E$ in each disorder realization, then its distribution has width of order one in either phase as $L\to \infty$ and $\xi_{FS}$ is not defined. 
If, however, we choose $X$ to be the level spacing ratio averaged over a sub-extensive energy window $\Delta E$ in each disorder realization, then we expect the distribution narrows as $L\to \infty$ and $\xi_{FS}$ can be defined. 

In the tractable examples of transitions between localized phases and delocalized non-ETH phases, $[r]$ is not a good order parameter for the transition. 
In the non-interacting Anderson model, the many-body levels always exhibit Poisson statistics and $[r]\approx 0.386$. Just to be clear, the level statistics of \emph{single-particle level spacings} jumps at the transition. 
Similarly, in the Clifford model, the presence of strictly localized integrals of motion everywhere in the phase diagram also leads to Poisson-like statistics. 

\emph{Relationship to current studies---}
In the last few years, studies of the MBL-delocalized transition have taken two approaches: numerical diagonalization of small systems and phenomenological real-space renormalization groups. The latter studies have found correlation length exponents consistent with the Harris bound, although their relationship to mean microscopic observables is unknown \cite{Vosk:2014aa,Potter:2015aa,Zhao:2015aa}. 

Recent numerical studies in one dimensional spin chains have reported scaling exponents violating the Harris bound \cite{Kjall:2014aa,Luitz:2015fj} as measured by order parameters which satisfy the conditions of the Mean Theorem. 
This either signals the failure of the finite-size scaling hypothesis, or that the largest systems are still too small to be in the scaling regime. 
One can estimate a crude upper bound on the system size $L^*$ beyond which the observed scaling forms cannot hold. 
This follows as the Lemma provides an absolute lower bound on the width of the scaling window and not just on the exponents. 
Unfortunately, $L^* \approx 500-5000$ for both numerical studies depending on which order parameter is considered. 
It is not surprising that $L^*$ is so large as 1) the Lemma uses no physical information about the order parameter, 2) the observed exponent likely increases with $L$, and 3) even in the scaling regime, Eq.~\eqref{Eq:Lemma} is likely not saturated. 
Nevertheless, $L^*$ provides some measure of asymptotic system sizes.
We have estimated $L^*$ for the standard deviation $\sigma_s = \sqrt{[(s- [s])^2]}$, the mean $[r]$, and the entanglement entropy generated in a local quench $[\Delta S]/L$. 
All of these satisfy the conditions of the Lemma.

We note that the constant $\alpha$ in the Lemma is formally infinite for the box distributions used in the numerical studies. In order to make the above estimates, we replace these distributions by Gaussians of the same width. 
It would be interesting to redo the numerical studies with disorder distribution which optimize the bound.


\emph{Discussion---}
In this article, we have derived elementary bounds on the finite-size scaling behavior of several order parameters associated with eigenstate phase transitions. 
In particular, the Harris bound holds for $[s]$ and $[r]$ at the localized-ETH transition and a weaker bound holds at general delocalization transitions without further assumptions. Our results are complementary to the celebrated CCFS bound.

Our mean results straightforwardly generalize to higher cumulants of bounded order parameters including the oft-studied variance of the entanglement entropy. 
They also can be readily applied to eigenstate phase transitions within the localized phase between states possessing distinct long-range quantum orders \cite{Huse:2013aa, Khemani:2015fj}.
For example, the spin-glass order parameter and the spin-spin correlation function at half system size in the random transverse field Ising model both satisfy the mean theorem. 
In this case, the stronger Harris bound can be proven using the strong-disorder renormalization group treatment \cite{Fisher:1995oq}.


Finite-size scaling likely applies even if the delocalization transition is ``first-order'' \cite{Privman:1983aa}. At a first order transition, there are no divergent length scales in the thermodynamic limit. Nonetheless, divergent finite-size scaling lengths control the scaling windows of $[s]$ and $[r]$. These satisfy the Harris bound by the Mean Theorem. 

At the other extreme, there could be multiple diverging length scales visible in the order parameters. This scenario would lead to the breakdown of the finite-size scaling hypothesis. It would be interesting to place general constraints on the transition in this case.

\emph{Acknowledgments---}
We are very grateful to S.L. Sondhi and D. Huse for their insights and detailed comments on this manuscript. We would also like to thank J. Bardarson, V. Khemani, J. Kj\"all and R. Moessner for helpful discussions. We thank MPIPKS (Dresden), ICTP and Perimeter Institute for their hospitality during the course of the project. Research at Perimeter Institute is supported by the Government of Canada through Industry Canada and by the Province of Ontario through the Ministry of Research and Innovation. VO acknowledges support from the NSF DMR Grant No. 0955714 and CRL from NSF PHY-1520535.

\bibliography{papers}

\end{document}